\documentclass[reprint,pra,aps,showpacs]{revtex4-1}

\usepackage{graphicx}
\usepackage{amsmath, amsthm, amssymb}
\usepackage{xcolor}

\begin{document}

\author{Johannes Flo\ss}
\author{Ilya Sh. Averbukh}
\title{Molecular spinning by a chiral train of short laser pulses}
\affiliation{Department of Chemical Physics, The Weizmann Institute of Sciences, Rehovot 76100, ISRAEL}
\date{\today}

\begin{abstract}
We provide a detailed theoretical analysis of molecular rotational excitation by a chiral pulse train -- a sequence of linearly polarised pulses with the polarisation direction rotating from pulse to pulse by a controllable angle.
Molecular rotation with a preferential rotational sense (clockwise or counter-clockwise) can be excited by this scheme.
We show that the directionality of the rotation is caused by quantum interference of different excitation pathways.
The chiral pulse train is capable of selective excitation of molecular isotopologues and nuclear spin isomers in a mixture.
We demonstrate this using $^{14}$N$_2$ and $^{15}$N$_2$ as examples for isotopologues, and para- and ortho-nitrogen as examples for nuclear spin isomers.
\end{abstract}


\maketitle



\section{Introduction}

The control of rotational molecular dynamics by non-resonant strong laser fields has proven to be a powerful tool.
It allows creating ensembles of aligned~\cite{friedrich95,*friedrich95b,stapelfeldt03,averbukh01,roscapruna01,ohshima10}, oriented~\cite{rost92,vrakking97} or planarly confined molecules~\cite{karczmarek99,fleischer09,kitano09,khodorkovsky10,lapert09,hoque11}.
The proposed and realised applications are numerous, including control of chemical reactions~\cite{larsen99,stapelfeldt03}, high harmonic generation~\cite{itatani05,wagner07}, control of molecular collisions with atoms~\cite{tilford04} or surfaces~\cite{kuipers88,tenner91,greeley95,zare98,shreenivas10}, and deflection~\cite{stapelfeldt97,purcell09,gershnabel10,*gershnabel10a} of molecules by external fields.

An important challenge for strong field rotational control is selective excitation in a mixture of different molecular species.
Isotopologue selective control was demonstrated in~\cite{fleischer06,fleischer08,akagi12}, using constructive and destructive interference induced by a pair of delayed laser pulses~\cite{lee06}.
With a similar scheme also nuclear spin isomer selective excitation was achieved~\cite{renard04,fleischer07,fleischer08}.
More recently, isotopologue selective rotational excitation by periodic pulse trains has been demonstrated~\cite{zhdanovich12a} and connection of this scheme to the problem of Anderson localisation was shown~\cite{floss12}.

One current direction of strong field rotational control focuses on the excitation of molecular rotation with a preferred sense of the rotation.
In the ``optical centrifuge'' approach~\cite{karczmarek99,spanner01,villeneuve00,spanner01a}, the molecules are subject to two counter-rotating circularly polarised fields, which are linearly chirped with respect to each other.
The resulting interaction potential creates an accelerated rotating trap, bringing the molecules to a fast spinning state.
The alternative ``double pulse'' scheme reaches the same goal by using two properly timed linearly polarised pulses~\cite{fleischer09,kitano09}.
The first pulse induces molecular alignment.
When the alignment reaches its peak, the second pulse, whose polarisation is rotated by 45 degree with respect to the first one, is applied and induces the directed rotation.
More recently, an approach using a ``chiral pulse train'' was demonstrated~\cite{zhdanovich11}, in which a train of pulses is used, where the pulse polarisation is changed by a constant angle from pulse to pulse (see Fig.~\ref{scenario}).

In this article, we provide a detailed theoretical analysis of the rotational excitation by a ``chiral pulse train'' demonstrated in~\cite{zhdanovich11}.
A thorough description of the experimental procedure is presented in a companion article~\cite{bloomquist12}.

For the present article, the structure is as follows.
In Section~\ref{sec.model} we introduce the model for the laser-molecule-interaction.
Then, we consider excitation scenarios for two kinds of molecules.
The first one is N$_2$, representing a simple diatomic molecule which is well described by the standard model of a rigid rotor.
The second molecule we consider is O$_2$.
Unlike the nitrogen molecule, it has a non-zero electronic spin in its ground state, leading to a more complex structure of the rotational levels.
Our analytical and numerical results are presented in Section~\ref{sec.results}.
Here, we first show the results for the excitation of $^{14}$N$_2$ by a chiral train of equally strong pulses.
Next, we demonstrate the prospects of selective excitation of nuclear spin isomers and isotopologues by such trains.
Finally, the results for oxygen molecules interacting with a train of non-equal pulses are shown and compared with the experiment~\cite{zhdanovich11}.
In the last section, we summarise the results and conclude.



\section{Model and Numerical Treatment}
\label{sec.model}

\subsection{Model}

\begin{figure}
\includegraphics{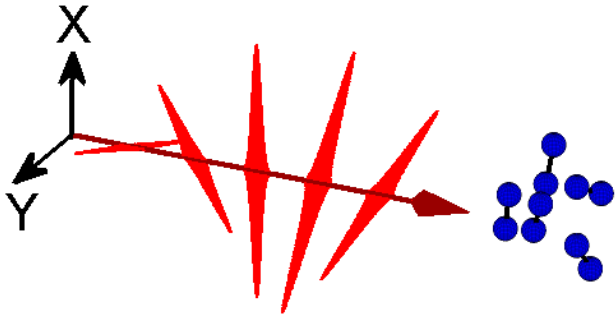}
\caption{\label{scenario}(Colour online) Sketch of the considered scenario:
A train of linearly polarised laser pulses interacts with linear molecules.
The polarisation axis is rotated by an angle $\delta$ between each pulse, and the time-delay $\tau$ between the pulses is constant.}
\end{figure}

We consider the following scenario:
A train of ultra-short laser pulses interacts with a gas sample of linear molecules like $N_2$ or $O_2$.
The pulses are applied with a constant delay $\tau$ between them.
Each pulse is linearly polarised, but the polarisation vector is rotated from one pulse to the next one by the angle $\delta$, such that the whole pulse train rotates with a rotational period $T_{train}=2\pi\tau/\delta$ (see Fig.~\ref{scenario}).
Choosing the laser propagation axis as $z$-axis, the electric field of the $n^{\text{th}}$ pulse is given as
\begin{equation}
E_n(t)=\mathcal{E}_n(t)\hat e_n \cos(\omega t + \phi) \,,
\label{eq.field}
\end{equation}
where $\hat e_n=\left(\cos n\delta, \sin n\delta, 0\right)$ is the polarisation vector, $\omega$ is the carrier frequency, and $\phi$ is the phase.
We consider a laser for which the carrier frequency is far detuned from any electronic or vibrational resonance.
The laser pulses therefore interact with the molecules via Raman-type excitations of the rotational levels.
The envelope of the electric field is given as
\begin{equation}
\mathcal{E}_n(t)=E_n\exp\left[-(t-n\tau)^2/(2\sigma^2)\right]\,.
\label{envelopeE}
\end{equation}
Here, $\sigma$ determines the pulse duration.

The non-resonant laser pulse induces a dipole in the molecule via its electric polarisability, and then interacts with this induced dipole.
Averaging over the fast oscillations of the electric field, we arrive at the effective interaction potential~\cite{boyd08}
\begin{equation}
V=-\frac{1}{4}\sum_n \mathcal{E}^2_n(t)\left(\Delta\alpha\cos^2\beta_n+\alpha_{\perp}\right)\,.
\label{potential}
\end{equation}
Here, $\Delta\alpha=\alpha_{\parallel}-\alpha_{\perp}$ is the polarisability anisotropy of the molecule, where $\alpha_{\parallel}$ and $\alpha_{\perp}$ are the polarisabilities along and perpendicular to the molecular axis, respectively.
The angle $\beta_n$ is the angle between the molecular axis and the polarisation direction of the $n^{\text{th}}$ pulse.
The last term in Eq.~\eqref{potential}, $\alpha_{\perp}$, is independent from the molecular orientation and does not influence the rotational dynamics.
We will therefore omit it in the following.

It is convenient to  introduce an effective pulse strength $P$, which corresponds to the typical amount of angular momentum (in units of $\hbar$) transferred to the molecule by the pulse.
For a single pulse, it is defined as
\begin{equation}
P=\frac{\Delta\alpha}{4\hbar}\int \mathrm{d}t\mathcal{E}^2(t)
=\frac{\Delta\alpha I_{peak}\sigma\sqrt{\pi}}{2c\epsilon_0\hbar}\,.
\end{equation}
Here, $I_{peak}$ is the peak intensity of the pulse, $c$ is the speed of light, and $\epsilon_0$ is the vacuum permittivity.

\begin{figure}
\includegraphics{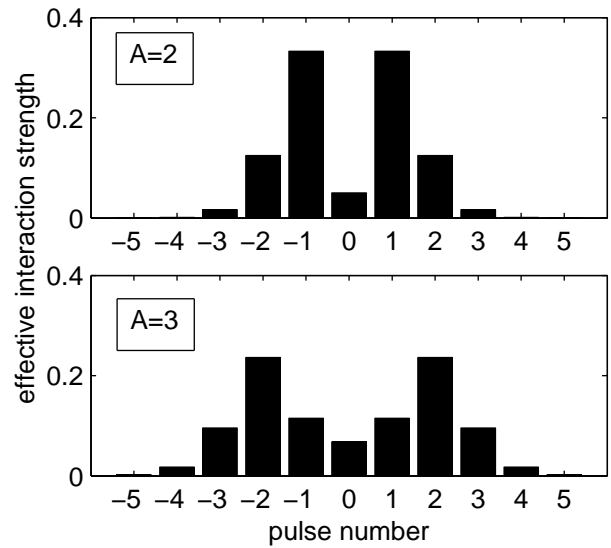}
\caption{\label{pulsetrain}Effective interaction strengths for a train with modulated intensities as described by Eq.~\eqref{envelopeBessel}. The total interaction strength is $P_{tot}=1$.}
\end{figure}

In this work, we consider two kinds of pulse trains.
The first one is a train of $N$ equally strong pulses, such that the effective interaction strength $P_n$ of the $n^{\text{th}}$ pulse is given as
\begin{equation}
P_n=P_{tot}/{N} \,,
\label{envelopeEqual}
\end{equation}
where $P_{tot}=\sum P_n$ is the total strength of the whole pulse train.
Such a pulse train can be generated e.g. by nested interferometers~\cite{siders98,cryan09}.
The second kind are trains like the ones used in the experiments~\cite{zhdanovich11,bloomquist12}, which were created by pulse shaping techniques.
In this case, the effective interaction strength of the $n^{\text{th}}$ pulse is given as
\begin{equation}
P_n=P_{tot} J_n^2(A) \,,
\label{envelopeBessel}
\end{equation}
where $J_n$ is the Bessel function of the first kind, and $A$ is a parameter.
Since $J_n(x)\ll1$ for $|n|>|x|$, this train contains about $2A+1$ non-zero pulses.
In Fig.~\ref{pulsetrain} we depict the intensity envelope of the train for different values of $A$.

\subsection{Numerical treatment}

\subsubsection{Nitrogen}

At first we consider molecular nitrogen as an example of a simple linear molecule.
Since the laser pulses are assumed to be far off-resonant from electronic or vibrational transitions, it is sufficient to consider only the rotational excitation in the vibronic ground state.
The rotational eigenfunctions are the spherical harmonics $|J,M\rangle$.
Here, $J$ is the total angular momentum, and $M$ is its projection on the $Z$-axis, which we have chosen to be along the laser propagation direction.
Note that for N$_2$ in its electronic ground state, the total angular momentum $\mathbf{J}$ is equal to the orbital angular momentum $\mathbf{O}$ of the rotation of the nuclei.
Although we are interested in the latter, for simplicity we keep to the more common notation using the total angular momentum $\mathbf{J}$.
The rotational levels are given as $E_J=BJ(J+1)-DJ^2(J+1)^2$, where $B$ is the rotational constant and $D$ is the centrifugal distortion constant.

For the numerical treatment of the problem it is convenient to express the wave function as a linear combination of the rotational eigenfunctions:
\begin{equation}
|\Psi(t)\rangle=\sum_{J,M} C_{J,M}(t) e^{-iE_Jt/\hbar} |J,M\rangle \,.
\label{wavepacket}
\end{equation}
Inserting the expansion~\eqref{wavepacket} and the interaction potential~\eqref{potential} into the time-dependent Schr\"odinger equation,
\begin{equation}
i\hbar\frac{\partial|\Psi(t)\rangle}{\partial t} = \hat H(t) |\Psi(t)\rangle \,,
\end{equation}
we obtain
\begin{multline}
i\hbar\sum_{J',M'} \frac{\partial C_{J'M'}(t)}{\partial t} e^{-iE_{J'}t/\hbar} |J',M'\rangle \\
=  \sum_{J',M'} C_{J'M'}(t) e^{-iE_{J'}t/\hbar} V(t) |J',M'\rangle \,.
\end{multline}
Multiplying from the left by $\frac{1}{i\hbar}\langle J,M|e^{iE_{J}t/\hbar}$ we obtain a set of coupled differential equations for the expansion coefficients $C_{J,M}(t)$:
\begin{align}
&\frac{\partial C_{JM}(t)}{\partial t} \nonumber\\
=&\frac{1}{i\hbar}\sum_{J',M'} C_{J'M'}(t) e^{-i(E_{J'}-E_J)t/\hbar} \langle J,M|V(t) |J',M'\rangle \nonumber\\
=&i\frac{\Delta\alpha}{4\hbar}\sum_{n=-\infty}^{+\infty}\mathcal{E}^2_n(t)\sum_{J',M'} C_{J'M'}(t) e^{-i(E_{J'}-E_J)t/\hbar} \nonumber\\
&\times\langle J,M|\cos^2\beta_n |J',M'\rangle \,.
\label{differentialEquation}
\end{align}
Here, $\beta_n$ is the angle between the molecular axis and the polarisation direction of the $n^{\text{th}}$ pulse.

The matrix element $\langle J,M|\cos^2\beta_n |J',M'\rangle$ is obtained as follows.
First, $\cos^2\beta_n$ is expressed as
\begin{multline}
\cos^2\beta_n = \cos^2(n\delta)\sin^2\theta\cos^2\phi+\sin^2(n\delta)\sin^2\theta\sin^2\phi\\
+\frac{1}{2}\sin(2n\delta)\sin^2\theta\sin(2\phi) \,,
\end{multline}
where $\theta$ and $\phi$ are the polar and azimuthal angle of the molecular axis, respectively.
Then, we express $\cos^2\beta_n$ in terms of the Wigner rotation matrices~\cite{brown03} $D_{MN}^{(J)}$ as
\begin{equation}
\cos^2\beta_n=\frac{1}{3}-\frac{1}{3}D_{00}^{(2)*}
+\frac{1}{\sqrt{6}}e^{i2n\delta}D_{-20}^{(2)*}+\frac{1}{\sqrt{6}}e^{-i2n\delta}D_{20}^{(2)*} \,.
\label{betaAsRotMat}
\end{equation}
Here, we use the relations
\begin{subequations}
\begin{multline}
\sin^2\theta\cos^2\phi=\\
\frac{1}{\sqrt{6}}\left(D_{20}^{(2)*}+D_{-20}^{(2)*}\right)-\frac{1}{3}D_{00}^{(2)*}+\frac{1}{3}
\end{multline}
\begin{multline}
\sin^2\theta\sin^2\phi=\\
-\frac{1}{\sqrt{6}}\left(D_{20}^{(2)*}+D_{-20}^{(2)*}\right)-\frac{1}{3}D_{00}^{(2)*}+\frac{1}{3}
\end{multline}
\begin{multline}
\sin^2\theta\sin(2\phi)=-i\sqrt{\frac{2}{3}}\left(D_{20}^{(2)*}-D_{-20}^{(2)*}\right)\,.
\end{multline}
\end{subequations}
Finally, by using~\cite{brown03}
\begin{multline}
\langle J,M |D_{M_00}^{(2)*}| J',M'\rangle
=(-1)^{M}\sqrt{(2J+1)(2J'+1)}\\
\times\left(\begin{array}{ccc}J&2&J'\\0&0&0\end{array}\right)\left(\begin{array}{ccc}J&2&J'\\-M&M_0&M'\end{array}\right)\,,
\label{couplingTerm}
\end{multline}
where the brackets denote the Wigner 3-j symbol, we obtain the matrix element $\langle J,M| \cos^2\beta |J',M'\rangle$.
Note that only levels with $\Delta J=0,\pm2,\pm4,...$ and $\Delta M=0, \pm2, \pm4, ...$ are coupled.

In our simulations, we solve Eq.~\eqref{differentialEquation} numerically.
We do ensemble averaging by solving Eq.~\eqref{differentialEquation} for different initial states $|\Psi_{initial}\rangle=|J_0,M_0\rangle$ and weighting the result by the Boltzmann factor of the initial state.
Note that the Boltzmann factor includes a degeneracy factor arising from nuclear spin statistics~\cite{herzberg89}.
For example, the nitrogen isotope $^{15}$N has a nuclear spin of $I=1/2$.
Therefore, the diatomic molecule $^{15}$N$_2$ can have a total nuclear spin of $I=1$ (ortho-nitrogen) or $I=0$ (para-nitrogen).
The former has three degenerate nuclear spin wave functions, which are symmetric with respect to an exchange of the two nuclei, and the later has one antisymmetric nuclear spin wave function.
Due to the fermionic nature of $^{15}$N, the total wave function of the molecule has to be antisymmetric with respect to the exchange of the nuclei.
Therefore, ortho- and para-nitrogen can be distinguished by their rotational wave functions:
Ortho-nitrogen is only found with odd angular momentum $J$, para-nitrogen only with even angular momentum $J$,
and the ratio of even to odd states is 1:3 due to the degeneracy of the nuclear spin wave functions of ortho-nitrogen.
For $^{14}$N with a nuclear spin of $I=1$, there are three nuclear spin isomers, two with symmetric nuclear spin wave functions (one of them five-fold degenerate), and one with three-fold degenerate antisymmetric nuclear spin wave functions.
The resulting ratio of even to odd rotational states is 2:1.

\subsubsection{Oxygen}

\begin{figure}
\includegraphics{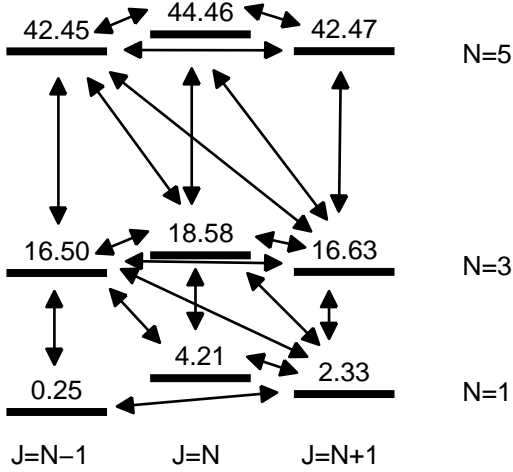}
\caption{\label{oxygenlevels}The lowest rotational levels of $^{16}$O$_2$ in its electronic and vibrational ground state.
The energies are given in $\text{cm}^{-1}$.
Also shown are the allowed transitions between these levels induced by the laser pulse~(\ref{potential}).}
\end{figure}

Molecular oxygen has a more complex rotational spectrum than simple diatomic molecules like nitrogen or hydrogen~\cite{townes55,brown03}.
The electronic ground state is a $^3\Sigma_g^-$ state, so the total electronic spin $\mathbf{S}$ is non-zero.
This gives rise to spin-spin and spin-orbit coupling, and therefore the total angular momentum $\mathbf{J}$ does not solely describe the nuclear rotational motion as for N$_2$, which has a $^1\Sigma_g^+$ electronic ground state.
In particular, $\mathbf{J}=\mathbf{N}+\mathbf{S}$, where $\mathbf{N}$ is the orbital angular momentum; since the electronic orbital angular momentum is zero, $\mathbf{N}$ is identical to $\mathbf{O}$, the nuclear orbital angular momentum.
The rotational quantum number $J$ can take the values $J=N+S,N+S-1,...,N-S$.
Therefore, for oxygen in its vibronic ground state with $S=1$, each level $N$ is split into three levels with $J=N-1,N,N+1$, as is shown in Fig.~\ref{oxygenlevels}.
The splitting is stronger for low values of $N$.
Additionally, for symmetry reasons, only odd values are allowed for $N$~\cite{brown03}.

For the numerical treatment, we express the wave function as a linear combination of Hund's case~(b) basis states~\cite{townes55,brown03}:
\begin{equation}
|\Psi\rangle = \sum_{JNM} C_{JNM}(t) e^{-i E_{JN} t/\hbar} |\eta \Lambda; N \Lambda; N S J M\rangle \,.
\label{wavepacketOxygen}
\end{equation}
Here, $\Lambda$ is the projection of the electronic angular momentum on the molecular axis, $N$ is the orbital angular momentum, $S$ is the electronic spin, $J$ is the total angular momentum, $M$ is the projection of the total angular momentum on the $Z$-axis, and $\eta$ is a combined quantum number of the remaining vibronic quantum numbers.
$E_{JN}$ are the energies of the rotational states, see Fig.~\ref{oxygenlevels}.
As before, we assume that the molecules are initially in their vibronic ground state.
Since the interaction does not induce any vibronic transitions, it is independent of $\eta$, and furthermore $\Lambda=0$ and $S=1$ are constant.
For ease of reading, in the following we denote the eigenstates in short as $|\eta \Lambda; N \Lambda; N S J M\rangle \equiv |JNM\rangle$.

As before, we insert the expanded wave function~\eqref{wavepacketOxygen} into the time-dependent Schr\"odinger Equation and obtain a system of differential equations for the expansion coeffecients $C_{JNM}$:
\begin{align}
\frac{\partial C_{JNM}(t)}{\partial t} &
=i\frac{\Delta\alpha}{4\hbar}\sum_{n=-\infty}^{+\infty}\mathcal{E}^2_n(t) \nonumber\\
&\quad\times\sum_{J',N',M'} C_{J'N'M'}(t) e^{-i(E_{J'N'}-E_{JN})t/\hbar} \nonumber\\
&\quad\times\langle J,N,M|\cos^2\beta_n |J',N',M'\rangle \,.
\label{differentialEquationOxygen}
\end{align}
The matrix elements $\langle JNM|\cos^2\beta|J'N'M'\rangle$ are determined as follows.
Firstly, we use Eq.~\eqref{betaAsRotMat} to replace $\cos^2\beta$, which yields
\begin{equation}
\begin{split}
\langle JNM|\cos^2\beta&|J'N'M'\rangle=\\
&\frac{1}{3}\langle JNM|J'N'M'\rangle\\
-&\frac{1}{3}\langle JNM|D_{00}^{(2)*}|J'N'M'\rangle\\
+&\frac{1}{\sqrt{6}}e^{i2n\delta}\langle JNM|D_{-20}^{(2)*}|J'N'M'\rangle\\
+&\frac{1}{\sqrt{6}}e^{-i2n\delta}\langle JNM|D_{20}^{(2)*}|J'N'M'\rangle \,.
\end{split}
\label{interactionterm}
\end{equation}
We now have to determine the value of $\langle \Lambda N S J M |D_{M_00}^{(2)*} | \Lambda N' S J' M'\rangle$.
Here we explicitly write down all quantum numbers (apart from $\eta$).
We use the Wigner-Eckart-theorem (see e.g.~\cite{brown03}) to exclude the dependence on the molecular orientation:
\begin{multline}
\langle \Lambda N S J M |D_{M_00}^{(2)*} | \Lambda N' S J' M'\rangle=(-1)^{J-M}\\
\times\left(\begin{array}{ccc}J&2&J'\\-M&M_0&M'\end{array}\right)\langle \Lambda N S J || D_{.0}^{(2)*}|| \Lambda N' S J'\rangle \,,
\end{multline}
where the round brackets is the Wigner 3-j symbol.
The dot in the subscript of the rotation matrix indicates that this matrix element is reduced regarding the orientation in the space-fixed coordinate system~\cite{brown03}.
Next, we use the fact that $D_{.0}^{(2)*}$ does not act on the electronic spin, so we can exclude $S$ from the matrix element as well and obtain~\cite{brown03}
\begin{multline}
\langle  \Lambda N S J || D_{.0}^{(2)*}|| \Lambda N' S J'\rangle=\\
 (-1)^{J'+N+2+S}\sqrt{(2J+1)(2J'+1)}\\
\times \left\{\begin{array}{ccc}N'&J'&S\\J&N&2\end{array}\right\} \langle  \Lambda N || D_{.0}^{(2)*}|| \Lambda N'\rangle \,.
\label{eqX}
\end{multline}
Here, the curly brackets denote the Wigner 6-j symbol.
Finally, the reduced matrix element in Eq.~(\ref{eqX}) is given as
\begin{multline}
\langle  \Lambda N || D_{.0}^{(2)*}|| \Lambda N'\rangle =\\
 (-1)^{N-\Lambda} \sqrt{(2N+1)(2N'+1)} \left(\begin{array}{ccc} N&2&N'\\-\Lambda&0&\Lambda \end{array}\right) \,.
\end{multline}
Here, we used Eq.~(5.186) in~\cite{brown03} and applied it for Hund's case (b).
We insert now $S=1$ and $\Lambda=0$, and obtain the matrix elements of the rotation matrices as
\begin{multline}
\langle \Lambda N S J M |D_{M_00}^{(2)*} | \Lambda N' S J' M'\rangle=\\
\times \sqrt{(2J+1)(2J'+1)(2N+1)(2N'+1)}\\
\times\left(\begin{array}{ccc}J&2&J'\\-M&M_0&M'\end{array}\right)
\left(\begin{array}{ccc} N&2&N'\\0&0&0 \end{array}\right)\\
\times\left\{\begin{array}{ccc}N'&J'&1\\J&N&2\end{array}\right\}
(-1)^{J+J'-M+1}
\label{interactionlongApp}
\end{multline}
Inserting Eq.~\eqref{interactionlongApp} into~\eqref{interactionterm} yields the matrix elements $\langle JNM|\cos^2\beta|J'N'M'\rangle$.
In order to lower the numerical effort, we treat the pulses as delta-pulses (sudden approximation), i.e. we neglect the molecular rotation during each pulse.
Comparison with experiment~\cite{zhdanovich11} shows that this approximation is well justified for pulses of a duration of 500~fs.
Using the method of an artificial time parameter $\xi$ as described in~\cite{fleischer09}, the differential equations for the expansion coefficients for a single laser pulse become
\begin{multline}
\frac{\partial C_{JNM}(\xi)}{\partial \xi} =iP_n\sum_{J',N'M'} C_{J'N'M'}(\xi) \\\times\langle J,N,M|\cos^2\beta_n |J',N',M'\rangle \,,
\label{differentialEquationOxygen2}
\end{multline}
where $P_n$ is the effective interaction strength introduced above.
Setting $C_{JNM}(\xi=0)$ to the values just before the pulse, we obtain the expansion coefficients right after the pulse as $C_{JNM}(\xi=1)$~\cite{fleischer09}.
To obtain the final expansion coefficients after the whole pulse train, we solve Eq.~\eqref{differentialEquationOxygen2} for every pulse, letting the wave packet~\eqref{wavepacketOxygen} evolve freely between the pulses.
To account for thermal effects, we do ensemble averaging over the initial state.
Since $^{16}$O has a nuclear spin of $I=0$, there are no degeneracies due to the nuclear spin wave functions.
However, only odd values are allowed for the orbital angular momentum $N$.



\section{Results}
\label{sec.results}

We will first present the results for excitation of nitrogen molecules by a train of equally strong pulses.
We will then demonstrate how such pulse trains can be used to selectively excite isotopologues and nuclear spin isomers in molecular mixtures.
Finally, we will show results for the excitation of the more complex oxygen molecules by a train of unequal pulses (given by Eq.~\eqref{envelopeBessel}), in order to compare our results with recent experiments~\cite{zhdanovich11, bloomquist12}.

We define the final population $Q(J)$ of a rotational level $J$ as
\begin{equation}
Q(J)=\sum_i g_i \sum_{M} |C_{i,JM}|^2\,.
\label{populationDef}
\end{equation}
Here, $i$ denotes the initial state and $g_i$ is its statistical weight.
We also define the directionality $\epsilon(J)$ of the excited wave packet as
\begin{equation}
\epsilon(J)=\frac{Q_L(J)-Q_R(J)}{Q_L(J)+Q_R(J)} \,,
\label{directionalityDef}
\end{equation}
where $Q_L(J)$ and $Q_R(J)$ are the counter-clockwise rotating and the clockwise rotating fraction of the population of the level $J$,
\begin{subequations}
\begin{align}
Q_L(J)&=\sum_i g_i \left(\sum_{M>0} |C_{i,JM}|^2+1/2|C_{i,J0}|^2\right)\\
Q_R(J)&=\sum_i g_i \left(\sum_{M<0} |C_{i,JM}|^2+1/2|C_{i,J0}|^2\right)\,.
\end{align}
\end{subequations}
Note that the states with $M=0$ are accounted half for clockwise and half for counter-clockwise rotation.
A positive (negative) $\epsilon(J)$ indicates a preferentially counter-clockwise (clockwise) rotation.


\subsection{Excitation of nitrogen molecules with a train of equally strong pulses}

In the following, we show the results for $^{14}$N$_2$ molecules interacting with a train of eight equally strong pulses with durations of $\sigma=30~\text{fs}$ (see Eq.~\eqref{envelopeE}) and a total interaction strength of $P_{tot}=5$.
The peak intensity of a single pulse is therefore approximately $3\cdot10^{12}\text{W}/\text{cm}^2$.
The pulse duration is well below the rotational periods of the highest expected excitations (remember that $P$ corresponds to the typical angular momentum in the units of $\hbar$ transferred by the pulse).
The molecules are considered being initially at a temperature of $T=8~\text{K}$.
At this temperature there is a considerable initial (thermal) population in the level $J=2$, with $Q_{th}(2)=0.25$.
Also in $J=3$ there is some initial population, $Q_{th}(3)=0.02$.
The levels $J=4$ and $J=5$ are not populated (note that due to nuclear spin statistics, two thirds of the population is found in the even levels, and one third in the odd ones).

\begin{figure}
\includegraphics{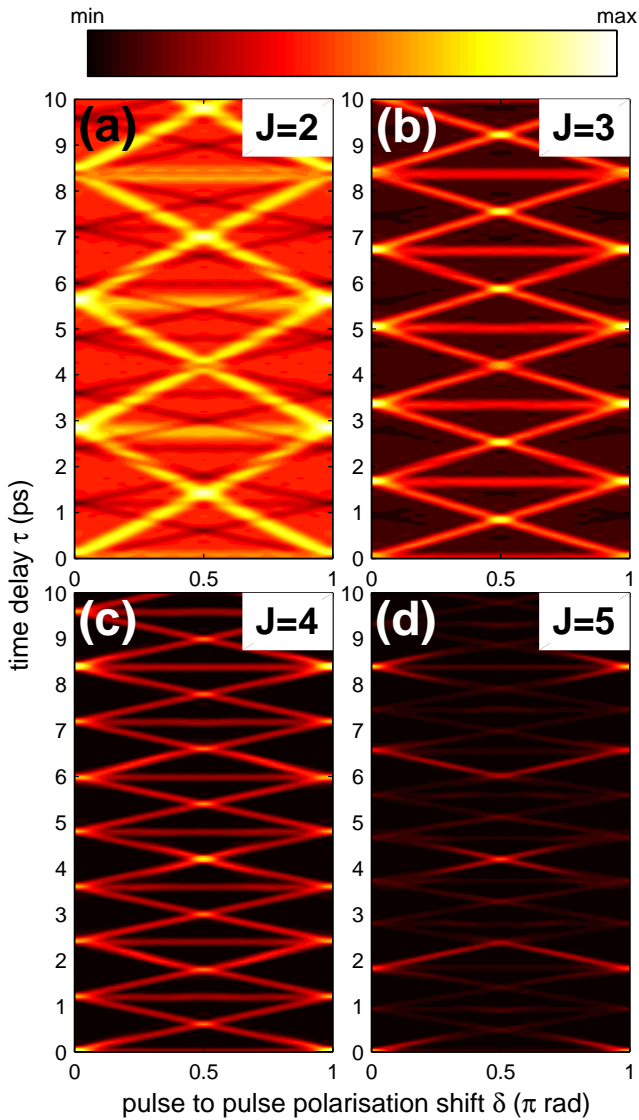}
\caption{\label{population_general}(Colour online) Population $Q(J)$ (see Eq.~\eqref{populationDef}) of different rotational levels $J$ for $^{14}$N$_2$ at $T=8 K$, after interacting with a train of eight equal pulses.
The total interaction strength is $P_{tot}=5$ and the pulse duration is $\sigma=30~\text{fs}$.
The minimum of the colour-coding is 0 for all panels, and the maximum is 0.6 for (a), 0.25 for (b) and (c), and 0.1 for (d).
}
\end{figure}

\begin{figure}
\includegraphics{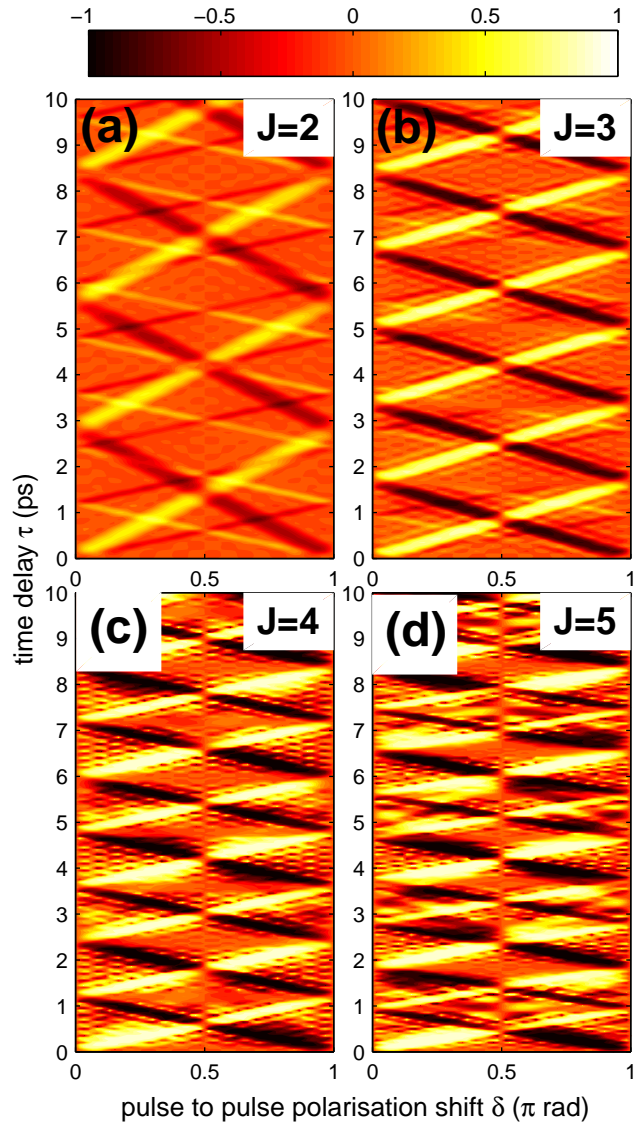}
\caption{\label{directionality_general}(Colour online) Directionality $\epsilon(J)$ (see Eq.~\eqref{directionalityDef}) of different rotational levels $J$ for $^{14}$N$_2$ at $T=8 K$, after interacting with a train of eight equal pulses.
The total interaction strength is $P_{tot}=5$ and the pulse duration is $\sigma=30~\text{fs}$.
Positive $\epsilon$ corresponds to counter-clockwise rotation, negative $\epsilon$ corresponds to clockwise rotation.}
\end{figure}

In Figure~\ref{population_general} the population $Q(J)$ is shown for the rotational levels $J=2,3,4, \text{and}~5$ for $^{14}$N$_2$ molecules, and Figure~\ref{directionality_general} displays the directionality $\epsilon(J)$ for the same levels.
The plots show the population and the directionality as a function of the pulse train period $\tau$ and the pulse-to-pulse polarisation angle shift $\delta$.

The population plots for all levels show a distinct pattern of diagonal and horizontal lines.
These lines are described by the equation
\begin{equation}
\tau= t_{exc}(J) \left(m+ \Delta M \frac{\delta}{2\pi}\right)\,.
\label{resonance_condition}
\end{equation}
Here, $m$ is an integer and $\Delta M=0,\pm2$ ($0$ yields the horizontal lines, $+2$ corresponds to the diagonal lines with a positive slope, and $-2$ yields the diagonal lines with a negative slope).
Furthermore, $t_{exc}$ is the period corresponding to the excitation from the level $J-2$ to the level $J$, and is given as
\begin{equation}
t_{exc}(J)=2\pi\hbar/(E_J-E_{J-2})=t_{rev}/(2J-1) \,,
\end{equation}
where $t_{rev}=\hbar\pi/B$ is the rotational revival time (8.38~ps for $^{14}$N$_2$ in its vibronic ground state).

The directionality plots in Fig.~\ref{directionality_general} show in general the same structure as the population plots, although now the horizontal lines are missing.
Furthermore, we can see that the diagonals with a positive slope correspond to a counter-clockwise rotational sense ($\epsilon(J)>0$), and the diagonals with a negative slope to a clockwise rotation.
Also, next to the main diagonals, there is a chessboard pattern visible, especially for higher levels.

The general structure of the population and directionality plots is very similar to experimental observations (Figs.~5 and~6 in~\cite{bloomquist12}), although the latter do not resolve the fine chessboard pattern.
It should be noted that in the experiments~\cite{bloomquist12} a train of unequal pulses described by~\eqref{envelopeBessel} was used, whereas the results presented here are for a train of equally strong pulses.

The structures seen in Figures~\ref{population_general} and~\ref{directionality_general} can be explained as the result of the quantum interference of different excitation pathways, as we will show now.
For simplicity, we will treat the pulses as delta-pulses in the following analysis.
This is well justified, as the utilised pulse duration of $\sigma=30~\text{fs}$ is much shorter than the relevant rotational periods of $^{14}$N$_2$.
The evolution of the wave packet over one period of the pulse train is given by
\begin{equation}
|\Psi(t_n^+)\rangle=e^{iP\cos^2\beta_n} e^{-i \hat J^2\tau/(2I\hbar)} |\Psi(t_{n-1}^+)\rangle \,,
\end{equation}
where $t_n^+$ is the time-instant right after the $n^{\text{th}}$ pulse, $\hat J$ is the angular momentum operator, and $I$ is the moment of inertia.
The interaction term can be expressed as
\begin{multline}
e^{iP\cos^2\beta_n} 
= \hat R(n\delta,\pi/2,0) e^{iP\cos^2\theta} \hat R^{-1}(0,-\pi/2,-n\delta) \\
= e^{-in\delta\hat J_z/\hbar} \underbrace{e^{-i\hat J_y\pi/(2\hbar)} e^{iP\cos^2\theta} e^{i\hat J_y\pi/(2\hbar)}}_{\hat W} e^{in\delta\hat J_z/\hbar} \,.
\label{interactionRotate}
\end{multline}
Here, $\hat R^{-1}$ rotates the basis from the space fixed system (quantisation axis along the laser propagation) to a ``pulse fixed'' system (quantisation axis along the electric field polarisation of the $n^{\text{th}}$ pulse).
The operator $\hat W$ is the same for every pulse.
Using~\eqref{interactionRotate}, we can express the evolution operator that brings the system from its initial state to the final state after the last pulse as
\begin{equation}
\hat U =\prod_{n=1}^N e^{-in\delta\hat J_z/\hbar} \hat W e^{in\delta\hat J_z/\hbar} e^{-i\frac{\hat J^2}{2I\hbar}\tau} \,.
\end{equation}
The probability of the transition from $|J'M'\rangle$ to $|JM\rangle$ is given as $\left|\langle JM|\hat U |J'M'\rangle\right|^2$.
Using the expansion
\begin{equation}
e^{iP\cos^2\theta}=1+iP\cos^2\theta-\frac{P^2}{2}\cos^4\theta+\ldots
\end{equation}
we can express the evolution operator as
\begin{align}
\hat U &=\prod_{n=1}^N e^{-in\delta\hat J_z/\hbar} e^{-i\hat J_y\pi/(2\hbar)}\left(1+iP\cos^2\theta+\ldots\right) \nonumber\\
&\qquad\times e^{i\hat J_y\pi/(2\hbar)} e^{in\delta\hat J_z/\hbar} e^{-i\frac{\hat J^2}{2I\hbar}\tau}\nonumber\\
&\approx e^{-i\frac{\hat J^2}{2I\hbar}\tau N} \nonumber\\
&\quad+ iP\sum_{n=1}^N e^{-i\frac{\hat J^2}{2I\hbar}\tau(N-n)} e^{-in\delta\hat J_z/\hbar} \nonumber\\
&\qquad \times e^{-i\hat J_y\pi/(2\hbar)} \cos^2\theta e^{i\hat J_y\pi/(2\hbar)} e^{in\delta\hat J_z/\hbar} e^{-i\frac{\hat J^2}{2I\hbar}\tau n} \,.
\label{eq.firstorder}
\end{align}
The approximation in the last line is valid in the limit of weak pulses ($P<1$).
In the following, we will only consider this limit.
Using Eq.~\eqref{eq.firstorder} as evolution operator, the total probability for a transition from the state $|J'M'\rangle$ to another state $|JM\rangle$ is given as
\begin{widetext}
\begin{align}
\left|\langle JM|\hat U |J'M'\rangle\right|^2
&=P^2\left|\sum_{n=1}^Ne^{-i(N-n)E_J\tau/\hbar}\langle JM|\hat V|J'M'\rangle e^{-i(M-M')n\delta}e^{-inE_{J'}\tau/\hbar}\right|^2 \nonumber\\
&=P^2|\langle JM\left|\hat V\right|J'M'\rangle|^2
\underbrace{\sum_{n,n'=1}^{N} \cos\left[\left(\frac{\Delta E \tau}{\hbar}-\Delta M \delta\right) (n-n') \right]}_{\equiv \Phi} \,.
\label{eq.phi}
\end{align}
\end{widetext}
Here, $\hat V=e^{-i\hat J_y\pi/(2\hbar)} \cos^2(\theta) e^{i\hat J_y\pi/(2\hbar)}$ and $\Delta E = E_J-E_{J'}$.
The term $\Phi$, and therefore the transition amplitude, is maximised, if the first factor in the argument of the cosine is a multiple of $2\pi$, which yields
\begin{equation}
\tau= \frac{2\pi\hbar}{\Delta E} \left(m+ \Delta M \frac{\delta}{2\pi}\right)\,,
\label{resonance_condition2}
\end{equation}
where $m$ is an integer.
This condition is equivalent to~\eqref{resonance_condition} and exactly describes the lines in Figures~\ref{population_general} and~\ref{directionality_general}.

Using these insights, we can now explain the results seen in Fig.~\ref{population_general} and~\ref{directionality_general}.
The patterns are the result of quantum interferences of different excitation pathways.
These interferences are constructive, when the condition~\eqref{resonance_condition2} is fulfilled, causing the lines seen in Fig.~\ref{population_general} and~\ref{directionality_general}.
By the help of Eq.~\eqref{resonance_condition2} we can also see that the horizontal lines are due to transitions with no change of the projection $M$, i.e. $\Delta M=0$.
There are no horizontal lines in the directionality plots, since $\Delta M=0$ means that there is no change in the sense of the rotation.
The diagonals with a positive slope are due to transitions with $\Delta M=2$.
The increase of $M$ shifts the rotational sense towards a counter-clockwise direction, and therefore increases the directionality $\epsilon(J)$.
The opposite is found for the diagonals with a negative slope, which correspond to $\Delta M=-2$.

\begin{figure}
\includegraphics{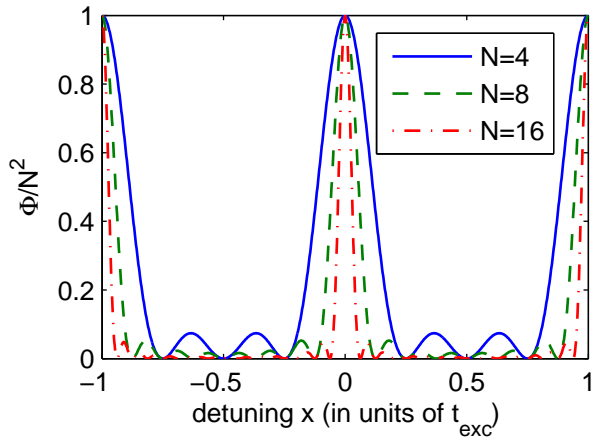}
\caption{\label{detuning}(Colour online) Sum $|\Phi|$ of the interference terms in Eq.~\eqref{eq.phi} as a function of the detuning $x$ from the condition~\eqref{resonance_condition2}.
The plot shows $\Phi$ for three different numbers $N$ of pulses, and the result is normalised to $N^2$.
}
\end{figure}

The chessboard pattern seen in the directionality plot can be explained by looking at the sum $\Phi$, when $\tau$ is detuned from the condition~\eqref{resonance_condition2}:
\begin{equation}
\tau= t_{exc} \Delta M \frac{\delta}{2\pi} + x\,.
\end{equation}
Here, $t_{exc}=2\pi\hbar/\Delta E$.
In Figure~\ref{detuning} we plot $\Phi$ as a function of the detuning $x$.
It can be seen that next to the main peaks at integer $x/t_{exc}$, there are weak oscillatory beats in between.
The minima of those beats are found at $x/t_{exc}=m/N$, where $N$ is the number of pulses and $m$ and $N$ are mutually prime.
These ``side bands'' are weak and therefore they can not be seen in the population plots in Fig.~\ref{population_general}.
On the other hand, the directionality measures the relative difference of the populations, so these weak side bands become visible, if the thermal population of the rotational level is sufficiently small.

\begin{figure}
\includegraphics{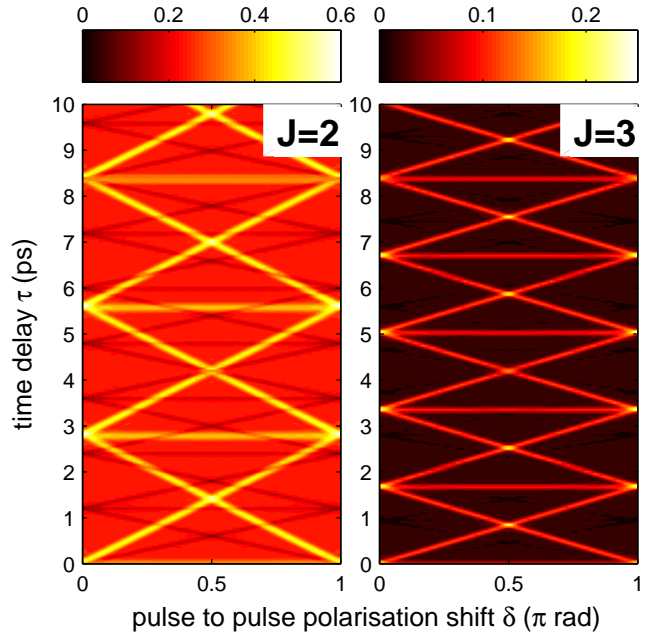}
\caption{\label{Ndependence}(Colour online) Population $Q(J)$ (see Eq.~\eqref{populationDef}) of the rotational levels $J=2$ and $J=3$ for $^{14}$N$_2$ at $T=8 K$, after interacting with a train of 16 equal pulses.
The total interaction strength is $P_{tot}=5$ and the pulse duration is $\sigma=30~\text{fs}$.}
\end{figure}

From Figure~\ref{detuning} we can see that an increase of the number of pulses leads not only to an increase in the number of ``side bands'', but also to a narrowing of the main peak of $\Phi(x)$.
Therefore, we expect a narrowing of the lines seen in the population plots for larger $N$.
This can be seen in Fig.~\ref{Ndependence} for the population of the levels $J=2$ and $J=3$.
Here, we use the same parameter values as in Fig.~\ref{population_general}, but twice as many pulses, while keeping the total interaction strength $P_{tot}$ constant.
A similar effect of the narrowing of the resonance when increasing the number of pulses was already found for the quantum resonance at the full rotational revival~\cite{floss12}.


\subsection{Selective Excitation}

\subsubsection{Nuclear spin isomer selective excitation}

The nitrogen isotopologue $^{15}$N$_2$ can be found as ortho-nitrogen with a total nuclear spin of $I=1$, or as para-nitrogen with a total nuclear spin of $I=0$.
These spin isomers can be distinguished by their rotational wave functions~\cite{herzberg89}:
Ortho-nitrogen is only found with odd angular momentum $J$, para-nitrogen only with even angular momentum $J$.

\begin{figure}
\includegraphics{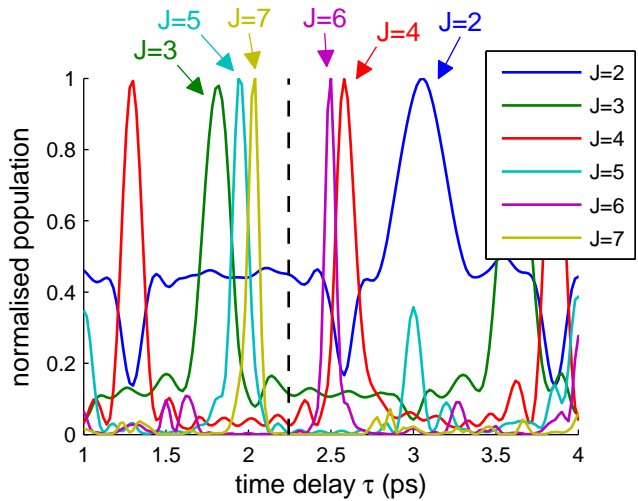}
\caption{\label{state_selective}(Colour online) Final population of the rotational levels of $^{15}$N$_2$ after interaction with 8 equal pulses with a duration of $\sigma=30~\text{fs}$ and a non-rotating polarisation ($\delta=0$) and $P_{tot}=5$, at $T=8~\text{K}$.
Each curve is normalised to its maximum.
The dashed line indicates the quarter revival time $t_{rev}/4$.
For $\tau$ slightly smaller than $t_{rev}/4$, only odd states are excited, for $\tau$ slightly larger, only even states.}
\end{figure}

\begin{figure}
\includegraphics{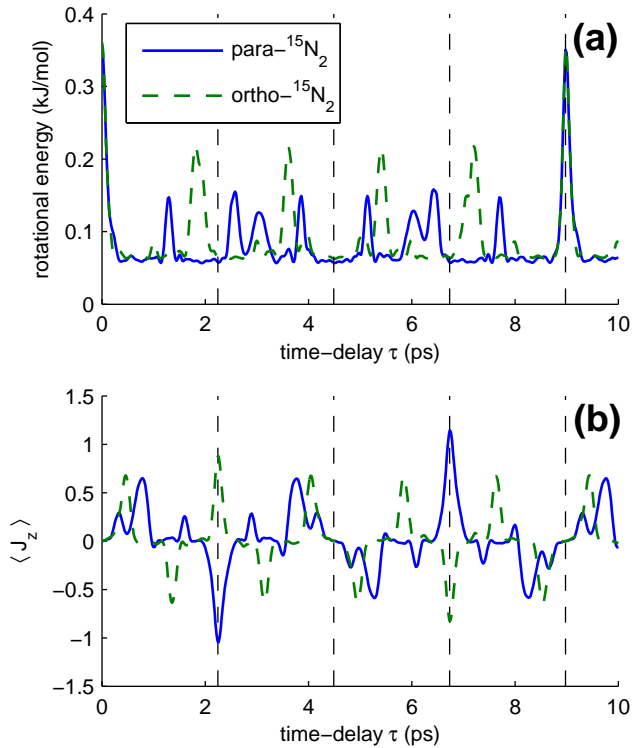}
\caption{\label{isomer}(Colour online) Selective excitation of the nuclear spin isomers of $^{15}$N$_2$ by interaction with a periodic pulse train of eight equally strong pulses with a duration of $\sigma=30~\text{fs}$.
The molecules are initially at $T=8~\text{K}$.
The total interaction strength is $P_{tot}=5$.
The results are shown as a function of the pulse train period $\tau$.
(a) Absorbed rotational energy for a non-rotating ($\delta=0$) pulse train.
(b) Projection of the molecular angular momentum on $Z$ for a rotating ($\delta=\pi/4$) pulse train.
The angular momentum is given in units of $\hbar$.
The dashed vertical lines indicate quarter, half, three quarter and full rotational revival times.
}
\end{figure}

In Figure~\ref{state_selective} we show the final population for the lowest rotational levels, after excitation by a pulse train with $\delta=0$ and period $\tau$ close to one quarter of the revival time (marked by the dashed line).
By tuning the time-delay between the pulses one can choose which state is excited strongest.
Moreover, for $\tau<t_{rev}/4$ only odd levels are significantly excited, whereas for $\tau>t_{rev}/4$ only even levels are significantly excited.
Therefore, by choosing $\tau$ slightly smaller (slightly larger) $t_{rev}/4$, one selectively excites ortho-nitrogen (para-nitrogen).
This effect is shown in Figure~\ref{isomer}~(a), which displays the absorbed energy for both spin isomers.
One can see separate peaks for both isomers, in particular close to $\tau=t_{rev}/4$ and $\tau=3t_{rev}/4$.
This selective excitation of spin isomers around a quarter of the revival time was demonstrated in a recent experiment~\cite{zhdanovich12a}.

At $\delta\neq0$, one may use chiral pulse trains to bring different spin isomers to a rotation of opposite sense.
The best selectivity is achieved for $\delta=\pi/4$.
In particular, at $\delta=\pi/4$ and $\tau=t_{rev}/4$ all excited even states have a positive directionality, and all excited odd states have a negative directionality.
The reverse is found at $\tau=3t_{rev}/4$.
The opposite directionality of even and odd rotational states at $\tau=t_{rev}/4$ and $\tau=3t_{rev}/4$ was also demonstrated experimentally (see Fig.~7 in~\cite{bloomquist12}).
This effect allows for spin isomer selective excitation, as shown in Fig.~\ref{isomer}~(b).
Here, the projection of the angular momentum on the $z$-axis is shown for both spin isomers.
For $\tau=t_{rev}/4$, ortho-nitrogen exhibits counter-clockwise rotation, whereas para-nitrogen rotates clockwise.
The opposite is found at $\tau=3t_{rev}/4$.

\subsubsection{Isotopologue selective excitation}

\begin{figure}
\includegraphics{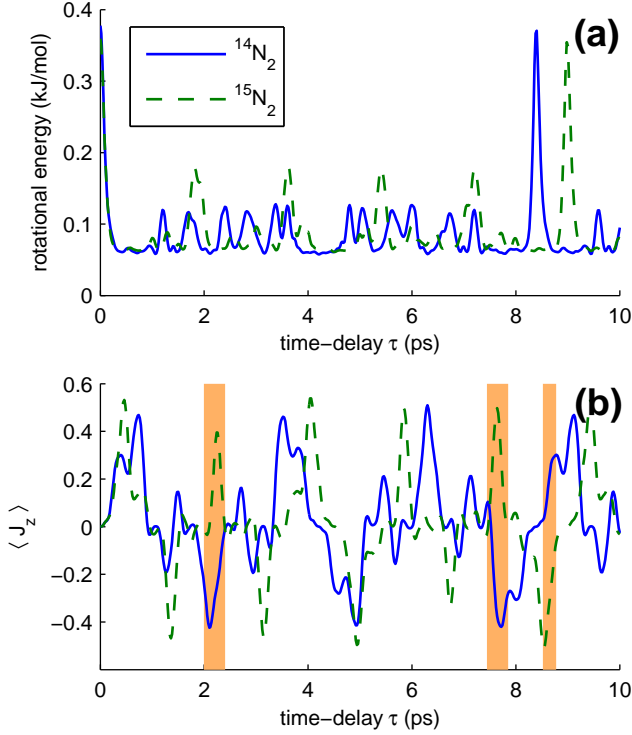}
\caption{\label{isotope_selective}(Colour online) Selective excitation of the nitrogen isotopologues $^{14}$N$_2$ and $^{15}$N$_2$ by interaction with a periodic pulse train of eight equally strong pulses with a duration of $\sigma=30~\text{fs}$.
The molecules are initially at $T=8~\text{K}$.
The total interaction strength is $P_{tot}=5$.
The results are shown as a function of the pulse train period $\tau$.
(a) Absorbed rotational energy for a non-rotating ($\delta=0$) pulse train.
(b) Projection of the molecular angular momentum on $Z$ for a rotating ($\delta=\pi/4$) pulse train.
The angular momentum is given in units of $\hbar$.
The shaded areas mark time-delays for which the two isotopologues rotate in opposite direction after the excitation.
}
\end{figure}

Isotopologues are chemically identical molecules with a different isotopic composition, e.g. $^{14}$N$_2$ and $^{15}$N$_2$.
Due to the different moments of inertia, isotopologues have different rotational time-scales.
For example, $^{14}$N$_2$ has a rotational revival time of $t_{rev}=8.38~\text{ps}$, whereas for $^{15}$N$_2$ the revival time is $t_{rev}=8.98~\text{ps}$.
Using that the rotational excitation is strongest if the pulse train period $\tau$ equals the rotational revival time, we can selectively excite $^{14}$N$_2$ or $^{15}$N$_2$ by tuning the train period to $\tau=8.38~\text{ps}$ or $\tau=8.98~\text{ps}$, respectively~\cite{floss12,bloomquist12}.
In Figure~\ref{isotope_selective}~(a) we show the absorbed rotational energy of the two nitrogen isotopologues after interaction with a non-rotating ($\delta=0$) pulse train, and one can clearly see the selective excitation at the respective revival times.

Inducing counter-rotation of different isotopologues is more challenging.
For spin isomers the time-scales were identical, and one could use the different directionality of even and odd states for $\tau=t_{rev}/4$ to induce counter-rotation.
For isotopologues the time-scales are different.
Counter-rotation can only be excited if for some set of parameters the pulse train accidentally excites rotation of opposite direction in the isotopologues.
For $^{14}$N$_2$ and $^{15}$N$_2$ three such regions can be seen in Fig.~\ref{isotope_selective}~(b) (see shaded region):
At $\tau\approx2.2~\text{ps}$ and $\tau\approx7.65~\text{ps}$ the heavier isotopologue rotates predominantly counter-clockwise ($\langle J_z \rangle >0$), and the lighter isotopologue rotates mainly clockwise; at $\tau\approx8.65~\text{ps}$ the opposite is found.


\subsection{Oxygen molecules in a chiral pulse train}

As a special example, we now consider the excitation of oxygen molecules by a chiral pulse train.
Instead of identical pulses we consider the more complex pulse sequence~\eqref{envelopeBessel}, corresponding to the one used in experiment~\cite{zhdanovich11} (see Fig.~\ref{pulsetrain}).
We also use parameters corresponding to this experiment:
The total interaction strength is $P_{tot}=7.5$, and $A=2$ (see Eq.~\eqref{envelopeBessel}).
Therefore, the strongest pulse in the train has an effective interaction strength of $P=2.5$, which corresponds to a peak intensity of approximately $8\cdot10^{12}\text{W}/\text{cm}^2$.
Due to the higher numerical complexity of the problem, we only considered delta-pulses.
Comparison with experiments~\cite{zhdanovich11} shows that this approximation is well justified.

Unlike molecular nitrogen, oxygen has a non-zero total electronic spin in its ground state.
There is a coupling between the electronic spin and the orbital angular momentum, leading to splitting of the rotational levels as shown in
Figure~\ref{oxygenlevels}.
Also, the orbital angular momentum $N$ is not identical to the total angular momentum $J$ any more, but $J=N-1,N,N+1$.
Note that due to the symmetry of the molecule, only odd values are permitted for $N$.

For oxygen, we define the population $Q(N)$ of a rotational level $N$ as
\begin{equation}
Q(N)=\sum_i g_i \sum_{J=N-1}^{N+1}\sum_{M=-J}^{J} |c_{i,JNM}|^2\,.
\label{populationDefOxygen}
\end{equation}
Here, $i$ denotes the initial state and $g_i$ is the corresponding statistical weight.
The population of counter-clockwise rotating states $Q_L(N)$ and clockwise rotating states $Q_R(N)$ is given as
\begin{subequations}
\begin{align}
Q_L(N)&=\sum_i g_i \left(\sum_{J,M>0} |c_{i,JNM}|^2+1/2|c_{i,JN0}|^2\right)\\
Q_R(N)&=\sum_i g_i \left(\sum_{J,M<0} |c_{i,JNM}|^2+1/2|c_{i,JN0}|^2\right)\,.
\end{align}
\end{subequations}
The directionality of the states with given $N$ is defined as
\begin{equation}
\epsilon(N)=\frac{Q_L(N)-Q_R(N)}{Q_L(N)+Q_R(N)} \,.
\label{directionalityDefOxygen}
\end{equation}

\begin{figure}
\includegraphics{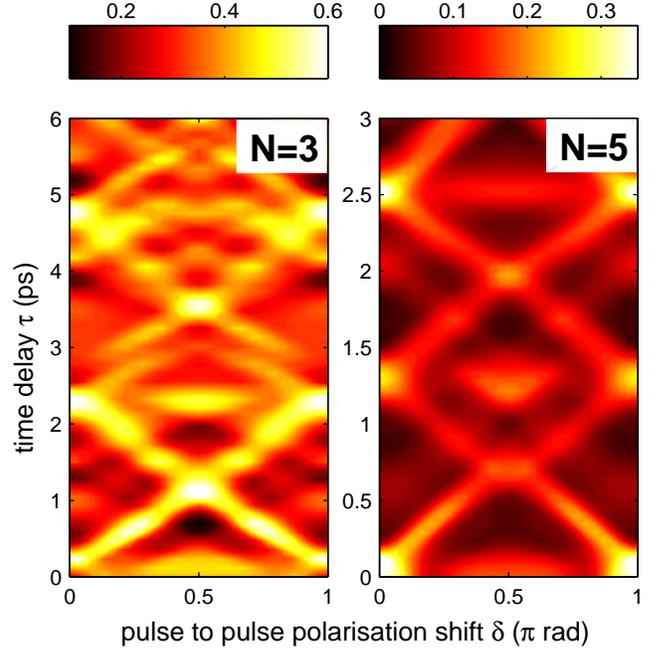}
\caption{\label{population}(Colour online) Population $Q(N)$ (see Eq.~\eqref{populationDefOxygen}) of the rotational levels $N=3$ and $N=5$ for $^{16}$O$_2$ at $T=8 K$, after interacting with a train of delta-pulses with intensity envelope given by~\eqref{envelopeBessel} ($A=2$).
The total interaction strength is $P_{tot}=7.5$.
Note the different scales for the ordinates.}
\end{figure}

\begin{figure}
\includegraphics{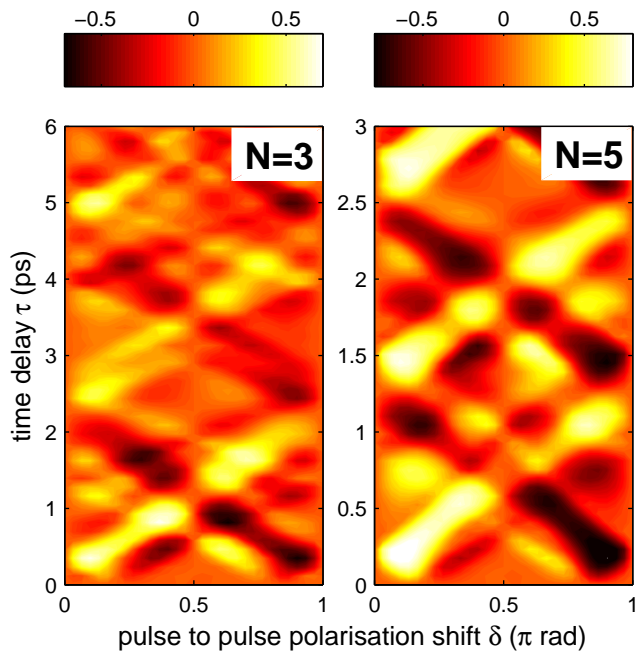}
\caption{\label{directionality}(Colour online) Directionality $\epsilon(N)$ (see Eq.~\eqref{directionalityDefOxygen}) of the rotational levels $N=3$ and $N=5$ for $^{16}$O$_2$ at $T=8 K$, after interacting with a train of delta-pulses with intensity envelope given by~\eqref{envelopeBessel} ($A=2$).
The total interaction strength is $P_{tot}=7.5$.
Note the different scales for the ordinates.}
\end{figure}

In Figure~\ref{population} we show the final population of the rotational levels $N=3$ and $N=5$ after the interaction with the pulse train.
In Figure~\ref{directionality} the directionality of these levels is shown.
We can see the same basic line structure as before for the nitrogen molecules.
However, due to the different pulse train and because of the level splitting, the lines are broader, and the pattern becomes more complex, especially for larger values of $\tau$.
On the other hand, the chessboard pattern seen in the directionality plots for molecular nitrogen is not found for molecular oxygen.
This is due to the fact that the chessboard pattern is caused by very weak side bands, which do not exist for molecular oxygen due to the level splitting.
One can also see that the plots for the level $N=3$ look more complex than the plots for the level $N=5$.
This is caused by the relatively stronger splitting for lower rotational levels.
Our calculated results very well resemble the ones from experiment~\cite{zhdanovich11}.



\section{Conclusions}


In this paper, we provided a detailed theoretical analysis of molecular rotational excitation by a ``chiral pulse train'', which was introduced in~\cite{zhdanovich11} and presented in more detail in the accompanying experimental paper~\cite{bloomquist12}.
The chiral pulse train is formed by linearly polarised pulses, uniformly separated in time by a time-delay $\tau$, with a constant pulse-to-pulse angular shift $\delta$ of the polarisation direction.
We showed that for certain combinations of $\tau$ and $\delta$, molecular rotation with a strong preferential rotational sense (clockwise or counter-clockwise) can be excited.
In two-dimensional plots of the excited population and the rotational directionality as a function of the pulse train period and the pulse-to-pulse polarisation shift, a distinct pattern is found.
It is made out of diagonal lines along which a strong preferential rotational sense is achieved.
Our analysis shows that this pattern is caused by quantum interferences of different excitation pathways, which interfere constructively along the above mentioned lines.

We demonstrated the feasibility for selective excitation of nuclear spin isomers and isotopologues in a mixture by the chiral pulse train.
We demonstrated the selectivity using para-nitrogen and ortho-nitrogen as an example.
By choosing the parameters of the chiral pulse train such that they address only the states of certain parity, one can selectively excite one of the isomers.
Since for the chiral pulse train one can also influence the direction of the molecular rotation, it is even possible to induce counter-rotation of different nuclear spin isomers.
Selective excitation of isotopologues can be reached by making use of the different rotational time-scales of different isotopologues.
The pulse train parameters can be chosen such that they lead to strong excitation of a preferable isotopologue.
For other isotopologues in the mixture, the same pulse train most likely leads to a destructive interference of different excitation pathways, so these isotopologues are at best only weakly excited.
Spin isomer and isotopologue selective excitation using the chiral pulse train was recently shown in experiments~\cite{zhdanovich12a,bloomquist12}, demonstrating a good agreement with our theoretical analysis.

Finally, we investigated the excitation of the more complex $^{16}$O$_2$ molecule by the chiral pulse train.
For this molecule, the rotational levels are split due to spin-spin and spin-orbit interactions.
We also used a slightly more complex pulse train as employed in experiment~\cite{zhdanovich11}.
In spite of these complications, our main conclusions remain valid also for the oxygen molecule.


We thank Erez Gershnabel, John Hepburn, Valery Milner and Sergey Zhdanovich for fruitful discussions.
Financial support of this research by the ISF (Grant No. 601/10) and the DFG (Grant No. LE 2138/2-1) is gratefully acknowledged.
The work of JF is supported by the Minerva Foundation.
IA is an incumbent of the Patricia Elman Bildner Chair.
This research is made possible in part by the historic generosity of the Harold Perlman Family.




%

\end{document}